\input harvmac
\def\half{{1 \over 2}}

\def\>{{\rangle}}
\def\<{{\langle}}

\def\p{{\partial}}

\def\L{{\Lambda}}
\def\vp{{\varphi}}

\def\a {{\alpha}}
\def\b {{\beta}}
\def\ad {{\dot a}}

\def\g {{\gamma}}
\def\d {{\delta}}

\def\e {{\epsilon}}

\Title{\vbox{\hbox{IFT-P.039/96}}}
{\vbox{\centerline{\bf 
Local Actions with Electric and Magnetic Sources}}}
\bigskip\centerline{Nathan Berkovits}
\bigskip\centerline{Instituto
de F\'{\i}sica Te\'orica, Univ. Estadual Paulista}
\centerline{Rua Pamplona 145, S\~ao Paulo, SP 01405-900, BRASIL}
\bigskip\centerline{e-mail: nberkovi@power.ift.unesp.br}
\vskip .2in
Superstring field theory was recently used to derive a covariant 
action for a self-dual five-form field strength. This action is 
shown to be a ten-dimensional version of the McClain-Wu-Yu action. 
By coupling to D-branes, it can be generalized in the presence
of sources. In four dimensions, this gives a local Maxwell 
action with electric and magnetic sources.

\Date{October 1996}
\newsec {Introduction}

Because duality symmetry relates theories with strong and weak coupling,
it is interesting to look for actions where this symmetry is manifest.
This problem is closely related to looking for actions of
self-dual $p$-form field strengths in $2p$ dimensions where $p$ is odd.
If manifest Lorentz invariance is sacrificed, it is straightforward
to construct quadratic actions for these systems.\ref\deser
{S. Deser and
C. Teitelboim, Phys. Rev. D13 (1976) 1592\semi
M. Henneaux and C. Teitelboim, Phys. Lett. B206 (1988) 650.}\foot{
However, it is not clear if these 
actions can be generalized in the
presence of sources.\ref\rohrlich{F. Rohrlich, Phys. Rev. 150 (1966) 1104.}
Note that the action of reference \ref\carneiro
{P.C.R. Cardoso de Mello, S. Carneiro, and M.C. Nemes,
Phys. Lett. B384 (1996) 197.} does not
seem to
give the correct Maxwell's equations of motion.
For example, varying $A^m$ gives
$\p^n F_{mn}(x)=-j_m(x) +h_m(x)$ where
$h_{m}(x)=\half\epsilon_{mnpq}\int d^4 y g^n(y) \int_{\tilde P}^y
\p^p \d^4(x-\xi) d\xi^q$ 
and ($j_m(x)$,$g_m(x)$) are the electric and magnetic sources. 
} Although manifest Lorentz invariance can
be recovered by introducing ``harmonic-like'' variables, the resulting
action is non-polynomial and may be difficult to quantize.\ref\sorokin{
P. Pasti, D. Sorokin, and M. Tonin, Phys. Rev. D52 (1995) R4277.}

An alternative approach is to use the Hamiltonian formalism. After
introducing an infinite set of fields, it is possible to construct
a set of first-class covariant constraints which impose the appropriate duality
conditions. 
This approach was first developed by McClain, Wu and Yu for 
two-dimensional chiral bosons\ref\MWY{B. McClain, Y.S. Wu, and
F. Yu, Nucl. Phys. B343 (1990) 689.}, and later generalized to four-dimensional
Maxwell 
\ref\martin{I. Martin and A. Restuccia, Phys. Lett. B323 (1994) 311.}
and self-dual $p$-forms\ref\devecchi{
F.P. Devecchi and M. Henneaux, ``Covariant path
integral for chiral p-forms'', hep-th 9603031.}.
Recently, it was shown by Bengtsson and Kleppe that by
performing a Legendre transformation, the McClain-Wu-Yu Hamiltonian
can be converted into a manifestly Lorentz-invariant action.\ref\bengtsson
{I. Bengtsson and A. Kleppe, ``On chiral P-forms'', hep-th 9609102.}  

Since the massless
Ramond-Ramond sector of the ten-dimensional Type IIB superstring
contains a self-dual five-form field strength, 
it is natural to ask how superstring field theory solves the problem.
Due to picture-changing difficulties, the massless Ramond-Ramond
contribution to the field theory action has only recently been computed in
\ref\me{N. Berkovits,
``Manifest electromagnetic duality in closed superstring field theory'',
hep-th 9607070, to appear in Phys. Lett. B.},
using the techniques of \ref\siegel
{W. Siegel, Int. J. Mod. Phys. A6 (1991) 3997\semi
N. Berkovits, M.T. Hatsuda, and W. Siegel, Nucl. Phys. B371
(1991) 434.} and
\ref\zwiebach{W. Siegel, Phys. Lett. B151 (1985) 396\semi
W. Siegel and B. Zwiebach, Nucl. Phys. B263 (1986)
105.}. 
Because of bosonic ghost zero modes, there
are an infinite number of fields in the superstring action, so one
suspects a relationship with the McClain-Wu-Yu action. Indeed, as
will be shown in this paper, the two actions coincide after gauge-fixing
certain fields.\foot{The fact that these two action coincide was also noticed
independently by Dmitri Sorokin.
\ref\sorokinp{D. Sorokin, private communication.}}

By coupling to D-branes, it is possible to generalize this action
in the presence of sources. Because
of manifest duality, electric and magnetic D-branes couple
symmetrically. 
After dimensional reduction to four dimensions,
this gives for the first time a local Maxwell
action with electric and magnetic sources.
Dirac charge quantization is implied by the existence of solutions
to the classical equations of motion.

In section II of this paper, the free superstring field theory
action of reference
\me is reviewed. In section III,
it is shown how this action reduces to the McClain-Wu-Yu action
after algebraically gauge-fixing certain fields.
In section IV, the D-brane boundary state is constructed,
and 
its contribution to the action
is computed. In section V, the action is generalized to four-dimensional
Maxwell in the presence of 
electric
and magnetic sources. 
In section VI,
some concluding remarks are made, including a conjecture that
the eleventh dimension of $M$-theory might be related to 
ghost degrees of freedom in superstring theory.

\newsec {Review of Action without Sources}

In reference \me, it was shown that by adding a non-minimal set of
variables to the usual RNS variables, the free action for the
Ramond-Ramond sector can be computed from a $\<\Phi|Q|\Phi\>$ action.
These non-minimal variables were first introduced in \siegel and 
consist of a left-moving
pair of conjugate bosons $(\tilde\g_L,\tilde\b_L)$ of weight
$(-\half,{3\over 2})$, a left-moving
pair of conjugate fermions $(\chi_L,u_L)$
of weight $(-\half,{3\over 2})$, and their right-moving
counterparts, $(\tilde\g_R,\tilde\b_R)$ and $(\chi_R,u_R)$.
The BRST operator is modified to
$Q_{new}=Q_{RNS}+ \int d\sigma
(u_L\tilde\g_L
+u_R\tilde\g_R)$, 
so using the standard ``quartet'' argument,
the new non-minimal fields do not affect the physical cohomology.
Like the $\psi^\mu$ matter fields and $(\g,\b)$ ghost fields, 
$(\tilde\g,\tilde\b)$ and $(\chi,u)$ are defined to be odd under
$G$-parity. 

The advantage of adding the non-minimal fields is that  
they allow consistent
boundary conditions for the bosonic zero modes, since        
the zero modes of $\g -i\tilde\g$ and $\b -i\tilde\b$
can be defined to annihilate
the incoming ground state while the zero modes of
$\g+i\tilde\g$ and $\b +i\tilde\b$ annihilate the outgoing
ground state.  
As shown in reference \siegel, 
this solves the picture-changing
problem associated with the Ramond sector of superstring field theory.  

Using the methods of \zwiebach, it was shown
in \me that the massless components of the incoming Ramond-Ramond string 
field can be described by 
\eqn\massl{|\vp\>= f^{\a\b} (x^\mu,\psi^\mu_L,\psi^\mu_R, u_L,
u_R, y)|L_{\a}\>|R_\b\> }
where $\a$ is a Weyl or anti-Weyl SO(9,1) spinor index depending
if it is a superscript or subscript, 
\eqn\ydef{y={i\over 2}[(\g_L+i\tilde\g_L)
(\b_R+i\tilde\b_R)
-(\g_R+i\tilde\g_R)
(\b_L+i\tilde\b_L)],}
$|L_\a\>|R_\b\>$ is annihilated by the zero modes of
$\g_L -i\tilde\g_L$,
$\b_L -i\tilde\b_L$,
$\g_R -i\tilde\g_R$,
$\b_R -i\tilde\b_R$, and
transforms under the zero modes of $\psi_L^\mu$ and $\psi_R^\mu$ as
$$\psi_L^\mu |L_\a\>|R_\b\>=\gamma^\mu_{\a\g}|L^\g\>|R_\b\>,\quad
\psi_R^\mu |L_\a\>|R_\b\>=\gamma^\mu_{\b\g}|L_\a\>|R^\g\>,$$
and
$f^{\a\b}$ is
a GSO-projected real function of the zero modes of $x^\mu$,
$\psi_L^\mu$,
$\psi_R^\mu$,
$u_L$, $u_R$, and $y$.
($f^{\a\b}$ can only depend on $\b$ and $\g$ in the above combination
since it must be an SU(1,1) singlet in the language of \zwiebach.)
The GSO projection implies that
for the Type IIB superstring, $f^{\a\b}$ must have even left-moving
and right-moving $G$-parity, while for the Type IIA superstring, 
$f^{\a\b}$ must have even left-moving and odd right-moving $G$-parity.  

Therefore, for Type IIB, 
$$|\vp\>=\sum_{n=0}^\infty {{y^{2n}}\over {(2n+1)!}}
(F_{(2n)}^{\a\b}
|L_\a\> |R_\b\>  +u_L E_{(2n)\a}^\b 
|L^\a\> |R_\b\> $$
\eqn\closex{ +u_R D_{(2n)\b}^\a 
|L_\a\> |R^\b\>  +u_L u_R  C_{(2n)\a\b}  
|L^\a\> |R^\b\> )}
$$+\sum_{n=0}^\infty {{y^{2n+1}}\over{(2n+2)!}}
(F_{(2n+1)\a\b}
|L^\a\> |R^\b\>  +u_L E_{(2n+1)\b}^\a
|L_\a\> |R^\b\> $$
$$ +u_R D_{(2n+1)\a}^\b 
|L^\a\> |R_\b\>  +u_L u_R  C_{(2n+1)}^{\a\b}  
|L_\a\> |R_\b\> ).$$
So at the massless level, an infinite number of fields are present
in the Ramond-Ramond sector of the Type II
superstring. 
(For the Type IIA superstring, the only difference in $|\vp\>$ is in
the position of the right-moving spinor index.)

In order to remove subtleties associated
with an infinite number of
fields, it will be assumed that at each point in spacetime, 
there exists an $N$ such that for all $n>N$,
$|F_{(n)}^{\a\b}|<
{1\over n}$,
$|E_{(n)\a}^{\b}|<
{1\over n}$,
$|D_{(n)\b}^{\a}|<
{1\over n}$, and
$|C_{(n)\a\b}|<
{1\over n}$.
This limiting procedure
differs from that of 
reference \me, and is similar to the restriction
of reference \bengtsson. It implies that the action and
energy of the system is finite and well-defined.

The action for these fields was calculated in \me from a 
$\<\Phi|Q|\Phi\>$ action and is
$${\cal S}=  
\sum_{n=0}^\infty [
2C_{(n)\a\b}\p_\mu\p^\mu(\p^{\a\g}E_{(n)\g}^\b +
(-1)^n\p^{\b\g}D^\a_{(n)\g})$$ 
$$-2F_{(n)}^{\a\b}((-1)^n\p_{\b\g}E_{(n)\a}^\g +\p_{\a\g}D^\g_{(n)\b}) $$
\eqn\finalc{
-(F^{\a\b}_{(n)}+\p^{\a\g}E_{(n)\g}^\b+(-1)^n\p^{\b\g}D_{(n)\g}^\a
-(-1)^n\p^{\a\g}\p^{\b\d}C_{(n)\g\d})}
$$(F_{(n+1)\a\b}+(-1)^n\p_{\b\kappa}D_{(n+1)\a}^\kappa
-\p_{\a\kappa}E^\kappa_{(n+1)\b}
+(-1)^n\p_{\a\kappa}\p_{\b\e}C_{(n+1)}^{\kappa\e})]$$
where it is understood that for odd $n$, the positions of all spinor 
indices are reversed (e.g. the first term for $n=1$ is
$2C_{(1)}^{\a\b} \p_\mu\p^\mu\p_{\a\g}E_{(1)\b}^\g$).

\newsec{Equivalence with McClain-Wu-Yu action}

Although \finalc looks complicated, it is easy to analyze because
of the gauge invariances:\foot
{These invariances are slightly different
from those of reference \me. In the language of \me, these invariances are
obtained from  
$$\d|\vp\>= Q^j\sum_{n=0}^\infty {{(it_L^k t_{Rk})^n}\over
{(n+2)!}}$$
$$
(-2t_{jL}(\L_{(n)\a}^\b+(-1)^n\p^{\b\g}\Theta_{(n)\a\g})|L^\a\> |R_\b\> 
-2t_{jR}(\Omega_{(n)\b}^\a+\p^{\a\g}\Xi_{(n)\g\b})|L_\a\> |R^\b\> $$
$$
-2t_{jL}u_R\Theta_{(n)\a\b}|L^\a\> |R^\b\>
-2t_{jR}u_L\Xi_{(n)\a\b}|L^\a\> |R^\b\>).$$}
\eqn\comp{\d C_{(n)\a\b}=\Theta_{(n)\a\b}+\Xi_{(n)\a\b},\quad 
\d D_{(n)\b}^\a=\Omega_{(n)\b}^\a-\p^{\a\g}(\Theta_{(n)\g\b}-\Xi_{(n)\g\b}),}
$$\d E_{(n)\a}^\b=\L_{(n)\a}^\b+(-1)^n 
\p^{\b\g}(\Theta_{(n)\a\g}-\Xi_{(n)\a\g}) ,$$
$$
\d F_{(n)}^{\a\b}=\p^{\a\g}\L_{(n)\g}^\b+
(-1)^n\p^{\b\g}\Omega_{(n)\g}^\a
+(-1)^n\p^{\a\d}\p^{\b\g}(\Theta_{(n)\d\g}+\Xi_{(n)\d\g}),$$
$$\d D_{(n+1)\a}^\b= 
-\L_{(n)\a}^\b ,\quad
\d E_{(n+1)\b}^\a=\Omega_{(n)\b}^\a,$$
$$
\d F_{(n+1)\a\b}=
-(-1)^n\p_{\b\g}\L_{(n)\a}^\g-\p_{\a\g}\Omega_{(n)\b}^\g.$$

Since the gauge transformations parameterized by $\Theta_{(n)\a\b}
+\Xi_{(n)\a\b}$,
$\L_{(n)\b}^\a$,
and $\Omega_{(n)\a}^\b$ act algebraically on
$C_{(n)\a\b}$, $D_{(n+1)\a}^\b$ and $E_{(n+1)\b}^\a$, they can
be used to gauge
\eqn\gaugechoice{C_{(n)\a\b}=D_{(n+1)\a}^\b=E_{(n+1)\b}^\a=0.}
In this gauge, the only non-zero fields are 
$D_{(0)\b}^\a$, $E_{(0)\a}^\b$, and $F_{(n)}^{\a\b}$, and the
action of \finalc simplifies to
\eqn\simple{{\cal S}= \int d^{10}x
[-2F_{(0)}^{\a\b}(\p_{\b\g}E_{(0)\a}^\g +\p_{\a\g}D^\g_{(0)\b})}
$$-F_{(1)\a\b}
(\p^{\a\g}E^\b_{(0)\g}
+\p^{\b\g}D_{(0)\g}^\a)
-
\sum_{n=0}^\infty 
(F_{(2n)}^{\a\b}
+F_{(2n+2)}^{\a\b})
F_{(2n+1)\a\b}].$$
Note that \simple is gauge invariant under 
$$\d D_{(0)\b}^\a=-\p^{\a\g}\lambda_{(0)\g\b},\quad
\d E_{(0)\a}^\b=\p^{\b\g}\lambda_{(0)\a\g},$$
which comes from $\Theta_{(0)\a\b}-\Xi_{(0)\a\b}$ in \comp.

Since 
$\Omega_{(n)\b}^\a$ and 
$\L_{(n)\a}^\b$ also transform 
$D_{(n)\b}^\a$ and 
$E_{(n)\a}^\b$ algebraically, one could have used them to
gauge 
$D_{(n)\b}^\a=
E_{(n)\a}^\b=0$ (as opposed to gauging 
$E_{(n+1)\b}^\a=
D_{(n+1)\a}^\b=0$). In this way, it would naively seem that
one could gauge 
$D_{(n)\b}^\a=
E_{(n)\a}^\b=0$ for all $n$, including $n=0$.
However, for large $n$, there is a problem with this gauge since
it does not preserve the property that 
$|E_{(n)\a}^\b|<{1\over n}$ for all $n>N$.
For example, suppose $D_{(N)\b}^\a=1$, and
$D_{(N+1)\a}^\b=
E_{(N+1)\b}^\a=0$. After using $\Omega_{(N)\b}^\a$ to gauge
$D_{(N)\b}^\a$ to zero, 
$E_{(N+1)\b}^\a$ will equal $-1$, which
does not satify
$|E_{(N+1)\a}^\b|<{1\over {N+1}}$.
This problem does not
occur if one instead uses
$\L_{(N-1)\b}^\a$ to gauge
$D_{(N)\b}^\a$ to zero.

The equations of motion of \simple are easily calculated to be
\eqn\eqpre{F_{(0)}^{\a\b}+\p^{\a\g}E_{(0)\g}^\b+\p^{\b\g}D^\a_{(0)\g}=
-F_{(2)}^{\a\b}=F_{(4)}^{\a\b}= -F_{(6)}^{\a\b}= ...,}
$$2(\p_{\b\g}E_{(0)\a}^\g+\p_{\a\g}D^\g_{(0)\b})=
-F_{(1)\a\b}=F_{(3)\a\b}= -F_{(5)\a\b}= ...,$$
$$2\p_{\b\g}F_{(0)}^{\a\b}
+\p^{\a\d}F_{(1)\d\g}=
2\p_{\b\g}F_{(0)}^{\b\a}
+\p^{\a\d}F_{(1)\g\d}=0.$$

If we assume that there exists an $N$ such that 
$|F_{(n)}^{\a\b}|<{1\over n}$ for $n>N$, 
then the only solutions to \eqpre satisfy
\eqn\eqten{F_{(0)}^{\a\b}+\p^{\a\g}E_{(0)\g}^\b+\p^{\b\g}D^\a_{(0)\g}=0,}
$$\p_{\b\g}E_{(0)\a}^\g+\p_{\a\g}D^\g_{(0)\b}=0,$$
$$\p_{\b\g}F_{(0)}^{\a\b}=
\p_{\b\g}F_{(0)}^{\b\a}=0,$$
$$F_{(2n+2)}^{\a\b}=F_{(2n+1)\a\b}=0.$$
Expanding $F_{(0)}^{\a\b}$ in vector notation, these equations are easily
seen to imply the Bianchi identities and equations of motion
for a 1-form, 3-form, and self-dual 5-form field strength.\me

The action for a self-dual 5-form field strength can be extracted
from \simple as
\eqn\five{{\cal S}= \int d^{10}x}
$$
[(-2F_{(0)pqrst}-F_{(1)pqrst})\p^{p}A^{qrst}
-
\sum_{n=0}^\infty 
(F_{(2n)pqrst}
+F_{(2n+2)pqrst})
F_{(2n+1)}^{pqrst}]$$
where 
$A^{qrst}=
(\gamma^{qrst})^\b_\a
(D_{(0)\b}^\a+E_{(0)\b}^\a)$ and
$$F_{(2n)pqrst}=
{1\over{120}}\epsilon_{pqrstuvwxy}
F_{(2n)}^{uvwxy}
=F_{(2n)}^{\a\b}
(\gamma^{pqrst})_{\a\b},$$ 
$$F_{(2n+1)pqrst}=
-{1\over{120}}
\epsilon_{pqrstuvwxy}
F_{(2n+1)}^{uvwxy}
=F_{(2n+1)\a\b}
(\gamma^{pqrst})^{\a\b}.$$
This action
will now be shown to be equivalent to the McClain-Wu-Yu action for
a self-dual 5-form field strength.

After performing a Legendre transformation, the McClain-Wu-Yu action
is:\bengtsson\foot{This action differs by an overall sign from equation 27 of
reference \bengtsson since we use a metric of signature $(+,-,...,-)$.}
\eqn\MWYaction{{\cal S_{MWY}}= \int d^{10}x \sum_{n=0}^\infty {(-1)^n }}
$$[{1\over 4} G_{(n)pqrst}G_{(n)}^{pqrst}+
\L_{(n+1)}^{pqrst}(G_{(n)pqrst}-G_{(n+1)pqrst})
-\L_{(n+1)}^{pqrst} \L_{(n+2)pqrst}]$$
where $G_{(n)}^{pqrst}={1\over{120}}\p^{[p} B_{(n)}^{qrst]}$ and
$\L_{(n+1)pqrst}={{(-1)^n}\over{120}} \epsilon_{pqrstuvwxy}\L_{(n+1)}^{uvwxy}$.
Since \MWYaction is invariant under the gauge transformations
$$\d B_{(n)}^{pqrs}=\Theta_{(n)}^{pqrs},\quad
\d B_{(n+1)}^{pqrs}=\Theta_{(n)}^{pqrs},$$
$$
\d \L_{(n+1)}^{pqrst}=-{1\over{240}}
(\p^{[p}\Theta_{(n)}^{qrst]}+(-1)^n\epsilon^{pqrstuvwxy}
\p_{u}\Theta_{(n)vwxy}),$$
one can algebraically gauge $B_{(n+1)}^{pqrs}=0$ for all $n$. In this 
gauge, \MWYaction becomes
\eqn\MWYsimple{{\cal S_{MWY}}= \int d^{10}x }
$$[{1\over 4} G_{(0)pqrst}G_{(0)}^{pqrst}+
\L_{(1)}^{pqrst}G_{(0)pqrst}-\sum_{n=0}^\infty (-1)^n
\L_{(n+1)}^{pqrst} \L_{(n+2)pqrst}].$$
So $\cal S_{MWY}$ is equal to \five after identifying
$$A_{pqrs}={1\over 2}B_{(0)pqrs},\quad
F_{(0)}^{pqrst}=-\L_{(1)}^{pqrst}-{1\over 4}
(G_{(0)}^{pqrst}- {1\over{120}}\epsilon^{pqrstuvwxy}G_{(0)uvwxy}),$$
$$F_{(2n+1)}^{pqrst}=-(-1)^n\L_{(2n+2)}^{pqrst},\quad
F_{(2n+2)}^{pqrst}=(-1)^n\L_{(2n+3)}^{pqrst}.$$

\newsec{D-Brane Coupling}

In superstring field theory, there are no perturbative states
which act as sources for Ramond-Ramond fields.
However, it was recently shown that D-branes
act as non-perturbative sources for Ramond-Ramond fields.\ref\polchinski
{J. Polchinski, Phys. Rev. Lett. 75 (1995) 4724.}
At linearized level, the coupling is simply $\<\vp|{\cal D}\>$
where $\<\vp|$ is the Ramond-Ramond string field and
$|{\cal D}\>$ is the boundary state of the D-brane.\ref\callan
{J. Polchinski and
Y. Cai, Nucl. Phys. B296 (1988) 91\semi
C. Callan, C. Lovelace, C.R. Nappi, and S.A. Yost, Nucl. Phys. B308
(1988) 221.}\ref\yost{S.A. Yost, Nucl. Phys. B321 (1989) 629.}

So to add source terms to the action of \finalc, one simply needs
to construct the D-brane boundary state and compute its
contribution. For a $(P+1)$-dimensional D-brane, one defines
$|{\cal D}_P\>$ by requiring that\callan
\eqn\Drequirements{(\p_L x^j -\p_R x^j)|{\cal D}_P\>=
(\p_L x^\mu +\p_R x^\mu)|{\cal D}_P\>=0,}
$$(\psi_L^j -i\eta\psi_R^j)|{\cal D}_P\>=
(\psi_L^\mu +i\eta \psi_R^\mu)|{\cal D}_P\>=0,$$
$$(c_L +c_R)|{\cal D}_P\>=
(b_L -b_R)|{\cal D}_P\>=
(\g_L -i\eta\g_R)|{\cal D}_P\>=
(\b_L -i\eta\b_R)|{\cal D}_P\>=0$$
$$(\tilde\g_L +i\eta\tilde\g_R)|{\cal D}_P\>=
(\tilde\b_L +i\eta\tilde\b_R)|{\cal D}_P\>=
(u_L +i\eta u_R)|{\cal D}_P\>=
(\xi_L +i\eta\xi_R)|{\cal D}_P\>=0$$
where $j=0$ to $P$, $\mu=P+1$ to 9, and $\eta=+1$ or $-1$
(depending if it is a D-brane or an anti-D-brane).
This implies that $(Q_L+Q_R)|{\cal D}_P\>=0$ where
$Q_R+Q_L$ includes the term $\int d\sigma (u_L\tilde\gamma_L+
u_R\tilde\gamma_R)$.

Since we are interested in the coupling to massless Ramond-Ramond
fields, only the zero mode dependence of $|{\cal D}\>$ needs to be constructed.
It is easy to check that the requirements of \Drequirements are satisfied by
\eqn\Dstate{|{\cal D}_P\>=-2i\mu_P e^{i\eta y} \prod_{\mu=P+1}^9
\d (x^\mu-f^\mu)
(u_L+i\eta u_R)(c_L+c_R)}
$$(\gamma^{0 ... P})_\b^\a
(|L_\a\>|R^\b\> +i\eta|L^\b\>|R_\a\>)$$
where $y$ is defined in \ydef and $\mu_P$ is a constant which
can be determined from a one-loop calculation. (We are considering
the Type IIB superstring, so $P$ is odd. For the Type IIA superstring,
$P$ is even and the last line of \Dstate is replaced with
$(\gamma^{0 ... P})^{\a\b}
|L_\a\>|R_\b\> +
i\eta(\gamma^{0 ... P})_{\a\b}
|L^\a\>|R^\b\>$.) 
Note that no sum over pictures is necessary in the definition of
$|{\cal D}\>$.\yost

One can now use the fact that 
$$\<R_\a| \<L_\b| \bar y^m 
y^n 
~(c_L+c_R)~ u_{0L} u_{0R}~ h(x)|L^\g\> |R^\d\>$$
\eqn\normc{= \d_{n,m}~ n! (n+1)! ~(-1)^n 
\d_\a^\d \d_\b^\g \int d^{10}x h(x),}
to compute $\<\vp|{\cal D}\>$. Plugging in the expression
of \closex for $\<\vp|$, one finds
\eqn\sourceaction{{\cal S}_P=\<\vp|{\cal D}_P\>=}
$$2\eta \mu_P \int_{x^\mu=f^\mu} d^{P+1}x^j (\gamma^{0 ... P})_\a^\b
\sum_{n=0}^\infty (-1)^n (E_{(2n)\b}^\a -D_{(2n)\b}^\a
+E_{(2n+1)\b}^\a +D_{(2n+1)\b}^\a).$$

So in the presence of D-branes, the massless Ramond-Ramond
contribution to the Type IIB superstring field
theory action is
$${\cal S} ={\cal S}_{free} +\sum_{i=0}^5 {\cal S}_{2i-1}$$
where ${\cal S}_{free}$ is defined in \finalc
and ${\cal S}_{2i-1}$ is defined in \sourceaction.

This action is still invariant under the gauge transformations of \comp,
and after gauge-fixing $C_{(n)\a\b}=D_{(n+1)\a}^\b=E_{(n+1)\a}^\b=0$,
it simplifies to 
\eqn\sourcesimple{{\cal S}=
 \int d^{10}x
[-2F_{(0)}^{\a\b}(\p_{\b\g}E_{(0)\a}^\g +\p_{\a\g}D^\g_{(0)\b}) }
$$-F_{(1)\a\b}
(\p^{\a\g}E^\b_{(0)\g}
+\p^{\b\g}D_{(0)\g}^\a)
-
\sum_{n=0}^\infty 
(F_{(2n)}^{\a\b}
+F_{(2n+2)}^{\a\b})
F_{(2n+1)\a\b}$$
$$+2\sum_{P}\eta \mu_P
\int_{x^\mu=f^\mu} d^{P+1}x^j (\gamma^{0 ... P})_\a^\b
(E_{(0)\b}^\a -D_{(0)\b}^\a)].$$

The equations of motion of \eqten are now modified to
\eqn\eqA{F_{(0)}^{\a\b}+\p^{\a\g}E_{(0)\g}^\b+\p^{\b\g}D^\a_{(0)\g}=0,}
$$\p_{\b\g}E_{(0)\a}^\g+\p_{\a\g}D^\g_{(0)\b}=0,$$  
\eqn\eqF{\p_{\b\g}F_{(0)}^{\a\b}=
\sum_P\eta\mu_P (\gamma^{0 ... P})^\a_\g \delta^{9-P}
(x^\mu-f^\mu),}
$$\p_{\b\g}F_{(0)}^{\b\a}
=-\sum_P\eta\mu_P (\gamma^{0 ... P})^\a_\g \delta^{9-P}
(x^\mu-f^\mu),$$
$$F_{(2n+2)}^{\a\b}=F_{(2n+1)\a\b}=0.$$

By plugging equations \eqA into equations \eqF, it naively appears that
the charges $\mu_P$ must vanish. However, this is not true since
$D_{(0)\b}^\a$ and
$E_{(0)\a}^\b$ are only single-valued up to the gauge transformation
$$\d D_{(0)\b}^\a=-\p^{\a\g}\lambda_{(0)\g\b},\quad
\d E_{(0)\a}^\b=\p^{\b\g}\lambda_{(0)\a\g}.$$
The equations of motion, when combined with single-valuedness of
gauge-invariant objects, imply (using the usual arguments \ref
\teit{R.I. Nepomechie, Phys. Rev. D31 (1985) 1921\semi
C. Teitelboim, Phys. Lett. B167 (1986) 63, 69.}) that
the charges satisfy the quantization condition that ${1\over{2\pi}}
\mu_P \mu_{6-P}$ is an
integer and $\mu_9=0$. In other words, if ${1\over{2\pi}}\mu_P \mu_{6-P}$ is not
an integer or $\mu_9$ is non-zero, the equations of motion have no solution.

\newsec{Generalization to Four-Dimensional Maxwell}

In the
absence of sources, the four-dimensional
Maxwell action was found in reference \me by performing
dimensional reduction on \finalc. After algebraically gauge-fixing
$C_{(n)ab}=D_{(n+1)a\dot b}=E_{(n+1)\dot a b}=0$ and separating out
the Maxwell field, the action is
\eqn\fouraction{{\cal S}_{free}=
 \int d^{4}x
[-F_{(0)}^{pq}(\p_p A_q -\half\epsilon_{pqrs}\p^r B^s) }
$$+\half F_{(1)}^{pq}(\p_p A_q +\half\epsilon_{pqrs}\p^r B^s) 
-{1\over 2}
\sum_{n=0}^\infty 
(F_{(2n)}^{pq}
+F_{(2n+2)}^{pq})
F_{(2n+1)pq}]$$
where 
$F^{pq}_{(n)}=2 Re(\sigma^{pq}_{ab} F^{ab}_{(n)})$, 
$A^p=4\sigma^p_{a \dot b} Re(D_{(0)}^{a\dot b}+E_{(0)}^{\dot b a})$,
and
$B^p=4\sigma^p_{a \dot b} Im(D_{(0)}^{a\dot b}+E_{(0)}^{\dot b a})$.
($a$ and $\dot a$ are two-component Weyl indices which come
from dimensionally reducing a sixteen-component SO(9,1) spinor.)
Note that \fouraction is manifestly invariant under the duality rotation
$$\d F_{(n)pq}=\half (-1)^n \epsilon_{pqrs} F^{rs}_{(n)},\quad
\d A_p=B_p,\quad \d B_p=-A_p.$$

To couple to $M$ dyons, 
one simply adds to \fouraction the source term
\eqn\foursource{{\cal S}_{source}=
\sum_{I=1}^M \int d\tau [m_I\sqrt {\dot y_I^p \dot y_{I p}} +\dot y_I^p
(e_I A_p (y_I(\tau))+
g_I B_p (y_I(\tau)))]}
where $y_I^p(\tau)$ is the worldline of the $I^{th}$ dyon, $\dot y^p=
{{dy^p}\over {d\tau}}$, and $(e_I,g_I)$ is
its electric and magnetic charge.

Using the requirement that $|F_{(n)}^{pq}|<{1\over n}$ for $n>N$,
it is easily verified that the equations of motion for 
${\cal S}_{free}+
{\cal S}_{source}$ are
\eqn\eqfour{F_{(0)}^{pq}=\p^{[p} A^{q]}=\epsilon^{pqrs} \p_r B_s,
\quad F_{(2n+1)}^{pq}=
F_{(2n+2)}^{pq}=0,}
$$\p_q F^{pq}_{(0)}=\sum_{I=1}^M e_I \int d\tau \dot y_I^p\d^4 (x-y_I(\tau)),$$
$$\half\epsilon^{pqrs}
\p_q F_{(0)rs}=\sum_{I=1}^M g_I \int d\tau\dot y_I^p \d^4 (x-y_I(\tau)),$$
$$m_I {{d^2 y^p}\over {d\tau^2}}
=e_I F^{pq}_{(0)} {{d y_q}\over {d\tau}}
+\half g_I \epsilon^{pqrs} F_{(0)rs} {{d y_q}\over {d\tau}},$$
where the scale
gauge $\dot y_I^p \dot y_{I p} =1$ has been
chosen.
Because
$\exp [i \oint_C dx^p ( e_I A_p
+ g_I  B_p)]$ must be single-valued for any closed contour $C$, 
these equations of motion only contain solutions if $(e_I,g_I)$ satisfy
the Dirac-Zwanziger quantization condition $e_I g_J -g_I e_J= 2\pi n_{IJ}$
for some integers $n_{IJ}$.

Note that one can replace the particle sources with fields, e.g.
\eqn\fields{{\cal S}_{source}=i\sum_{I=1}^M
\int d^4 x \bar\psi_I^\ad (\p_p -i e_I A_p -i g_I B_p) \sigma^p_{a\ad} \psi_I^a}
where $\psi_I^a$ are dyonic
spinors. This gives the appropriate Maxwell equations of motion
\eqn\Max{F_{(0)}^{pq}=\p^{[p} A^{q]}=\epsilon^{pqrs} \p_r B_s,
\quad F_{(2n+1)}^{pq}=
F_{(2n+2)}^{pq}=0,}
$$\p_q F^{pq}_{(0)}=
\sum_{I=1}^M e_I \bar\psi_I^\ad \sigma^p_{a\ad}\psi_I^a,\quad
\half\epsilon_{pqrs}\p^q F^{rs}_{(0)}=\sum_{I=1}^M g_I
\bar\psi_I^\ad \sigma^p_{a\ad}\psi_I^a,$$
$$(\p_p -i e_I A_p -i g_I B_p)\sigma^p_{a\ad}\psi_I^a=0.$$
These equations (when combined with the 
fact that $\exp [i \oint_C dx^p (e_I A_p
+ g_I B_p)]$ is single-valued for any closed contour $C$)      
imply that 
$${1\over{2\pi}}\sum_{J=1}^M (e_I g_J-g_I e_J) 
\int_V d^3 x \bar\psi_J^\ad \sigma^0_{a\ad} \psi_J^a$$
is an integer for any volume $V$.
Although this is untrue at the classical level even when
$(e_I,g_I)$ satisfy the Dirac-Zwanziger quantization condition,
it is true at the quantum level if 
$\int_V d^3 x \bar\psi_J^\ad \sigma^0_{a\ad} \psi_J^a$
is interpreted as the expectation value of
$${{\<\Phi|\int_V d^3 x \bar\psi_J^\ad \sigma^0_{a\ad} \psi_J^a|\Phi\>}\over{
\<\Phi|\Phi\>}}$$
where $|\Phi\>$ is any normalizable state. This is because 
$\bar\psi_J^\ad (x) \sigma^0_{a\ad} \psi_J^a(x)$ is an operator with
eigenvalues $\sum_{i=1}^N\delta^3(x-y_i)$, where $\Phi$ is a 
state constructed with
$N$ creation operators.\foot{I thank
Warren Siegel for pointing this out to me.}      

\newsec{Conclusions}

In this paper, D-branes were used to construct local actions with
electric and magnetic sources. These actions contain an infinite number of
fields and are closely related to the McClain-Wu-Yu action. It should
be straightforward to quantize them using the methods of \MWY\martin\devecchi
\bengtsson.

Since these action appear in superstring field theory, it is natural to
try to interpret the infinite set of fields as coming from some stringy
degree of freedom. In fact, by looking at equation \closex, it is clear that
this new degree of freedom is simply the $y$ variable, which is constructed
from an SU(1,1)-invariant combination of worldsheet ghosts. In this sense, 
the massless Ramond-Ramond string field $\vp$ can be interpreted as
living in eleven dimensions, where the eleventh dimension is parameterized
by $y$.

Recently, it was conjectured that there exists an eleven-dimensional
theory called $M$-theory which, after compactification on $S_1$, is dual
to Type IIA superstring theory.\ref\Mtheory{P. Townsend,
Phys. Lett. B350 (1995) 184\semi E. Witten, Nucl. Phys. B443 (1995) 85.}
$D_0$-branes are associated with
Kaluza-Klein states in $M$-theory and their Ramond-Ramond charge comes
from momentum in the compactified eleventh direction.
Furthermore, the radius of compactification of the eleventh dimension is
related to the expectation value of the dilaton in superstring theory. By
compactifying on $S_1/Z_2$ instead of $S_1$, $M$-theory can be related to the
heterotic superstring.\ref\horava{P. Horava and E. Witten,
Nucl. Phys. B460 (1996) 506.}

Is there a connection between $y$ and the eleventh dimension of $M$-theory?
Although this question will not be answered here, three pieces of favorable
evidence will be presented.

Firstly, the $y$ dependence of the $D_0$-brane boundary state is simply
$e^{\pm iy}$, where the $\pm$ depends if it is a $D_0$-brane (carrying
momentum $P_{11}=+1$) or an anti-$D_0$-brane (carrying $P_{11}=-1$).
At least naively, this suggests that $y=x_{11}.$

Secondly, $y$ is constructed from worldsheet ghosts, which couple to
worldsheet curvature through their background charge. Therefore, it is possible
that $y$ might be related to the expectation value of the dilaton, which also
couples to worldsheet curvature.

Thirdly, if the eleventh dimension of $M$-theory were related to worldsheet
ghost degrees of freedom, it might explain in a stringy manner why
different superstring theories can be unified. As was shown in \ref
\ust{N. Berkovits and C. Vafa, Mod. Phys. Lett. A9 (1993) 653.},
by mixing matter and ghost degrees of freedom, 
it is possible to relate string theories
with different numbers of worldsheet supersymmetries.
Perhaps duality symmetry can be understood as rotations which mix matter and
ghost degrees of freedom.

Clearly, more 
evidence needs to be gathered before this question can be answered.
Hopefully, the three pieces of evidence suggested in this conclusion will 
motivate others to investigate if the eleventh dimension of $M$-theory can
be related to worldsheet ghost degrees of freedom.

\vskip 20pt

{\bf Acknowledgements:} I would like to thank I. Bengtsson, 
S. Carneiro, C. Hull, S. Mandelstam,
H. Ooguri, S. Rangoolam, W. Siegel, D. Sorokin, and H. Verlinde 
for useful conversations. This
work was financially supported by the Brazilian 
CNPq and FAPESP.

\listrefs
\end